\documentclass{article}

\usepackage[english]{babel}

\usepackage[letterpaper,top=2cm,bottom=2cm,left=3cm,right=3cm,marginparwidth=1.75cm]{geometry}

\usepackage{amsmath}
\usepackage{graphicx}
\usepackage[colorlinks=true, allcolors=blue]{hyperref}
\usepackage{xcolor}
\usepackage{float}
\usepackage{booktabs}
\usepackage[round,authoryear]{natbib}

\title{Clinical Reasoning in the Age of AI: Longitudinal Cognition and Human-AI Collaboration}

\author{
  Irene Yi$^1$, Grace Brown$^{1,*}$, Sufian Aldogom$^{2,*}$, Nathan Roll$^1$, Eric J. Basile DO$^6$, \\
  Pamela M. Resnikoff MD MPH$^4$$^,$$^6$, Bianca Sanchez MD$^5$$^,$$^6$, Chirag Lodha DO$^5$$^,$$^6$, \\
  Isaac Gutterman$^3$, Oscar Schiff$^3$, Keira Salata$^3$, Benjamin Mujkic$^3$, Ammar Ahmed MD$^6$ \\
  \small $^1$Stanford University \\
  \small $^2$Massachusetts Institute of Technology \\
  \small $^3$Harvard University \\
  \small $^4$Scripps Mercy Hospital San Diego, Pulmonary Medicine \\
  \small $^5$University of South Florida Tampa General Hospital, Morsani College of Medicine\\
  \small $^6$Aurevia MD \\
  \small{$^*$Primary authors}
}

\begin{document}
\maketitle

\begin{abstract}
As physicians turn to AI-powered systems to help meet the dual demands of speed and care quality, they are met with hallucinations and sycophancy. Understanding how doctors reason through clinical problems in real-world settings is critical for design of effective AI reasoning systems. While recent advances in medical AI have emphasized performance benchmarks and diagnostic accuracy, comparatively little attention has been paid to the structure of clinicians’ reasoning processes as they unfold over time, e.g., how they interact with electronic health records and operate under conditions of uncertainty and constraint. This study provides a comprehensive, empirically-grounded account of clinical reasoning and its relationship to current AI-mediated workflows through a mixed-methods design that combines qualitative interviews with structured survey data.

Across both components, clinical reasoning is conceptualized as a multi-dimensional, dynamic process involving hypothesis generation and revision, longitudinal integration, trajectory tracking, uncertainty management, and partially implicit judgment. Interview data elicit detailed reasoning traces through case-based narratives, capturing how clinicians make decisions over time. Complementary survey data provide structured measures of these reasoning dimensions, alongside systematic characterization of how AI-generated clinical documentation tools are used in practice.

Findings indicate that current AI systems are primarily deployed for encounter-level tasks such as documentation and summarization, and only partially align with physicians’ underlying reasoning processes. In particular, AI-generated representations often omit temporal or interpretive structures central to clinical decision-making, while core aspects of reasoning, especially those spanning multiple encounters, remain largely implicit and physician-driven. By integrating fine-grained qualitative insights with broader quantitative patterns, this study offers a unified framework for understanding clinical reasoning as a context-sensitive, temporally extended process and identifies key mismatches between clinician cognition and current AI design. These results provide concrete directions for the development of AI systems that more effectively align with and augment real-world clinical reasoning.

\end{abstract}

\section{Introduction}
Clinical reasoning is the core cognitive process underlying medical decision-making \citep{Norman2005, Elstein1978}. Clinicians routinely interpret complex, evolving patient data and make decisions under uncertainty \citep{Polanyi1966, Dreyfus1986}. As such, clinical reasoning is recognized as a dynamic process involving both analytic and intuitive modes of cognition, with expert physicians integrating pattern recognition and reflective judgment under conditions of uncertainty \citep{Norman2005, Eva2005,Croskerry2009}. Because the most consequential parts of reasoning are often the least visible, systems built and evaluated on documented outputs risk optimizing for what is easy to measure rather than for how clinicians actually reason, leaving their true cognitive contribution unverified. This challenge has become increasingly salient with the rapid integration of artificial intelligence (AI) into clinical workflows, where the effectiveness of these systems depends not only on their technical performance but on how well they align with and support physician reasoning in practice as well. 

Recent advances in medical AI have demonstrated impressive capabilities in tasks such as diagnostic prediction or clinical documentation \citep{Esteva2017, Rajpurkar2017}, reflecting the broader emergence of machine learning as a major force in healthcare, where computational systems are increasingly used to detect patterns and augment clinical work \citep{Beam2018, Topol2019}. However, these systems are typically developed and evaluated using task-specific benchmarks that isolate narrow components of clinical work, creating a gap between benchmark performance and clinical usefulness, especially when systems are evaluated on isolated outputs rather than on their ability to support human reasoning \citep{DoshiVelez2017, Rudin2019, Tonekaboni2019}. As a result, they often fail to account for the broader structure of clinical reasoning, which extends beyond individual encounters and involves integrating information and uncertainty over time and coordinating multiple, partially implicit cognitive processes. This gap raises a fundamental question: what aspects of clinical reasoning are actually being supported by current AI systems, and which remain outside their scope? 

At the same time, there is growing recognition that clinical reasoning is not fully captured in formal representations such as clinical notes or structured data fields. Physicians routinely rely on tacit knowledge or heuristic judgments, as well as longitudinal understanding of patient trajectories that are only partially externalized in documentation. The increasing use of AI-generated clinical documentation tools further complicates this landscape, as these systems introduce new forms of representation that may reshape how reasoning is recorded or acted upon. Understanding how these representations relate to underlying cognitive processes is therefore critical for evaluating the role of AI in clinical practice. 

This study approaches clinical reasoning and AI use as parts of a single, integrated system. We examine how reasoning is performed in-situ and how it interacts with AI-mediated workflows. To accomplish this, we utilize an approach that combines qualitative, interview-based methods that elicit detailed reasoning traces with structured survey measures that enable systematic characterization across clinicians. This mixed-methods design allows us to capture both the fine-grained dynamics of reasoning as it unfolds over time and the broader patterns that shape its interaction with AI systems. The central aim of the study is twofold: first, to develop an empirically grounded account of clinical reasoning as a dynamic and partially implicit process; and second, to identify where and how current AI systems align with, or don't align with, this process. By situating AI use within the context of real-world clinical cognition, the study seeks to move beyond performance-based evaluations and toward a more integrated understanding of human–AI collaboration in medicine. 

We adopt a specific dynamic conceptualization of clinical reasoning: physicians reason across trajectories that unfold over multiple visits, continually reconstructing and reinterpreting prior information as presentations evolve. Two commitments follow from this view and organize the rest of the paper. First, cognition is not equivalent to documentation \citep{Rudin2019}, given that clinical notes are a compressed, selective externalization shaped by workflow (and often billing and time) constraints. As such, much of the reasoning (unresolved concerns, contextual impressions, etc.) remains internal even when a record appears complete. Second, because current AI tools operate largely on these documentary representations and treat encounters as discrete units, the central challenge for clinical AI may be representational rather than predictive. By this, we mean clinical AI may be greatly improved by aligning outputs with the uncertainty-laden and reconstructive structure of physician cognition rather than improving accuracy on isolated outputs. We return to both commitments when interpreting the survey indices and interview codes.

\section{Background}
\subsection{Clinical Reasoning}
When physicians reason about a medical problem, they do not simply retrieve facts or follow linear decision trees. Instead, they interpret incomplete information, generate and revise hypotheses, weigh competing explanations, integrate formal evidence with experiential judgment, and more. Early work on medical problem solving described clinical reasoning as an iterative process of hypothesis generation and data gathering, showing that physicians often move back and forth between patient cues and possible diagnoses rather than proceeding through a purely sequential analytic process \citep{Elstein1978, Kassirer1978}. Later work further emphasized that diagnostic performance depends on domain knowledge, pattern recognition, contextual cues, and the ability to recognize when initial impressions require revision \citep{Norman2005, NormanEva2010} in addition to these general hypothesis-driven strategies (i.e., the cycle of hypothesis generation, data gathering, and iterative refinement described above), suggesting that clinical reasoning is a dynamic process shaped by many factors. 

A key feature of real-world clinical reasoning is that much of it is tacit. Physicians often draw on forms of professional knowledge that are difficult to explicitly document, including intuition, practical judgment, contextual awareness, experience-based recognition of meaningful patterns, and more \citep{Dreyfus1986}. Polanyi’s concept of tacit knowledge is useful here because it explains how professionals may “know” more than they can fully articulate \cite{Polanyi1966}. In clinical settings, this means that the reasoning behind a decision may not be completely captured in the written note, the problem list, the billing code, or even the physician’s own retrospective explanation. Eraut’s work on professional learning similarly highlights how expertise develops through informal, situated practice rather than only through formal rules or explicit instruction \citep{Eraut2000}. This view is also consistent with Schön’s account of reflective practice, in which professional expertise emerges through situated judgment and reflection-in-action and not just rules alone \citep{Schon1983}. Together, these perspectives suggest that any attempt to evaluate clinician reasoning must account for both explicit reasoning processes and the less visible forms of judgment embedded in practice.

\subsection{EHR Systems and Documentation Fragmentation}
The electronic health record has become one of the primary sites where clinical reasoning is operationalized. We see this in artifacts such as the assessment-and-plan section of progress notes or structured order sets, where diagnostic impressions and management decisions are externalized for communication and reuse. However, EHR documentation does not necessarily represent the full structure of clinician cognition. Clinical notes must serve multiple purposes at once: supporting patient care, communicating with other clinicians, satisfying billing and legal requirements, and producing structured data for reuse. \cite{Rosenbloom2011} describe this as a tension between flexible narrative documentation and structured, computable data capture. Narrative notes may preserve nuance and clinical context, while structured fields may make information easier to retrieve. At the same time, EHR usability challenges can shape how clinicians document or search for information, and work on health IT usability has shown that poor interface design and workflow misalignment can create burdens that affect clinical performance and safety \citep{Ratwani2018}. Similarly, research on clinical decision support adoption has found that things like workflow fit and implementation, as well as perceived trust and usefulness are major factors influencing whether physicians actually use these systems \citep{Khairat2018}. Prior work on unintended consequences of health information technology also shows that clinical information systems can introduce new challenges when they are poorly aligned with clinical practice, such as hindering cognitive workflow and even safety \citep{Ash2004}. As a result, the EHR should not be treated as merely a neutral container of clinical reasoning.

\subsection{AI Systems}
Against this background, AI systems are increasingly being introduced into healthcare workflows, including for purposes of diagnosis, risk prediction, image interpretation, clinical documentation, and decision support. Much of the early benchmark literature in medical AI demonstrated impressive performance on narrowly defined tasks, such as dermatology image classification and chest X-ray interpretation \citep{Esteva2017, Rajpurkar2017}. These studies were important in showing that machine learning systems could match or exceed human-level performance on specific prediction tasks. However, such benchmarks often evaluate isolated outputs and not broader reasoning processes. A model that performs well on a classification benchmark may still fail to support the way physicians actually reason in complex or ambiguous clinical situations \citep{Rudin2019, Tonekaboni2019}.

More recent work has emphasized that successful clinical AI requires more than predictive accuracy. \cite{Sendak2020} argue that translating machine learning into healthcare requires attention instead to things like implementation, workflow integration, organizational readiness, clinical value, and more. In other words, the most effective AI systems are actually those that complement human expertise and supporting decision-making \citep{Bansal2021, Seeber2020}. This perspective is especially important for clinician-facing tools because their value depends heavily on how effectively they integrate into existing reasoning practices. Otherwise, it may increase cognitive burden rather than facilitate clinical reasoning.

Trust and explainability are therefore central themes in the clinical AI literature. Studies of explainable AI in healthcare show that clinicians do not want generic explanations of generated responses; they want explanations that are relevant to the clinical task and the patient context. \cite{Bussone2015} found that explanations can shape trust in clinical decision support, but trust is not automatically improved by adding more information. \cite{Tonekaboni2019} further show that clinicians value explanations that help them understand when and why a model’s output should be used, particularly in relation to clinical context and uncertainty. So, clinical AI systems should be evaluated not only by whether they produce correct answers, but by whether they support the forms of reasoning clinicians actually use.

\section{Study 1: Survey}
Study 1 consists of a survey constructed to examine how physicians reason through patient care over time and how well current AI documentation tools capture that reasoning. We surveyed 39 physicians across multiple specialties using Likert-scale and open-ended items that were aggregated into composite reasoning indices. The results show that clinical reasoning is largely longitudinal and only partially externalized in documentation, revealing a representational mismatch with existing AI systems and motivating the development of AI that supports cross-encounter, trajectory-based reasoning.

\subsection{Methods}
\subsubsection{Participants}
The survey dataset comprised responses from 39 physicians across multiple specialties and clinical settings, including emergency medicine, cardiology, internal medicine, family medicine, and more. Survey participants ranged from trainees and early-career physicians to clinicians with more than fifteen years of experience, allowing the survey to capture variation across both institutional settings and levels of clinical expertise. 

\subsubsection{Survey Design}
The survey component operationalizes clinical reasoning as a set of distinct, measurable components, such as how physicians form expectations, generate hypotheses, select information, and track patients over time, so that each component can be measured separately and compared across participants. The instrument is structured to separately measure baseline reasoning processes and AI-mediated workflows, allowing for independent characterization as well as analysis of their interaction. Survey items capture key dimensions of reasoning identified in both the literature and interview data, including prior expectation formation, hypothesis generation, information selection, and confidence in relevance judgments. Longitudinal reasoning is assessed through measures of reliance on prior encounters and frequency of referencing past records, while temporal reasoning is captured through trajectory tracking and monitoring of change over time.

Additional survey measures target absence reasoning, cross-encounter reinterpretation, and retrospective updating of clinical understanding. The instrument also assesses implicit versus explicit reasoning, such as the extent to which reasoning is consciously articulated, as well as reliance on intuition versus deliberate analysis. To examine how reasoning is externalized, the survey includes questions about documentation practices, including what proportion of reasoning is recorded in clinical notes and which aspects remain implicit. Reconstruction processes are measured through reported effort required to reconstruct patient state and reliance on memory versus documentation.

The AI-focused section of the survey captures frequency and purpose of AI use, particularly in relation to documentation and summarization. It also assesses the perceived quality of AI-generated representations, starting with their accuracy and completeness and also including their alignment with physicians’ reasoning processes. Additional items probe perceived information loss or cognitive effects of AI use, such as changes in workload or depth of reasoning. Participants are also asked to identify desired future capabilities of AI systems, with particular attention to support for longitudinal and cross-encounter reasoning.

The survey instrument was designed to operationalize clinical reasoning as a multidimensional and temporally extended process rather than a single diagnostic event. To accomplish this, the instrument incorporated both Likert-scale and open-ended questions targeting multiple dimensions of reasoning and workflow, including hypothesis generation, longitudinal reasoning, temporal tracking, retrospective reinterpretation, absence reasoning, selective attention, tacit and intuitive judgment, documentation practices, and AI-mediated documentation workflows. Most quantitative items were scored on five-point scales measuring frequency or importance, as well as degree of agreement with certain statements, allowing comparison across reasoning dimensions while preserving variation in physician practice patterns and cognitive strategies.

\subsubsection{Analysis}
Analysis proceeded in two stages. First, to move beyond isolated survey items, conceptually related measures were combined into composite indices covering longitudinal reasoning, tacit reasoning, documentation externalization, and perceived alignment between AI-generated notes and physician cognition. Each index is computed as the mean of its constituent items on the shared five-point scale. These indices were summarized using descriptive statistics (means and standard deviations) to characterize recurring structures in how physicians' clinical reasoning processes. Second, to test whether AI exposure was associated with different reasoning profiles, index and item scores were compared between physicians who reported using AI-generated notes and those who did not. Open-ended responses were then reviewed thematically and used to interpret the quantitative patterns, particularly around perceived information loss and desired future AI capabilities. The composite indices are defined below.

\paragraph{Longitudinal Reasoning Index}
The Longitudinal Reasoning Index was constructed to measure the extent to which physicians engage in temporally extended and reconstructive forms of reasoning. This index was calculated by averaging responses across five survey dimensions: importance of prior encounters, frequency of referencing prior encounters, extent of trend-based reasoning, reinterpretation of prior encounters, and degree of retrospective updating. Together, these measures capture whether physicians reason primarily within isolated encounters or whether they continuously integrate historical information and evolving trajectories over time. High scores on this index indicate that physicians routinely synthesize information across multiple visits, revise prior interpretations, and track changes over time, reinforcing the argument that much of real-world reasoning depends on reconstructing evolving patient trajectories rather than simply assigning static diagnostic labels.

\paragraph{Tacit Reasoning Index}
The Tacit Reasoning Index was designed to capture the extent to which clinical reasoning remains difficult to fully articulate. This index was calculated by averaging survey responses related to difficulty verbalizing reasoning and reliance on intuition during clinical decision-making. Higher scores on this index suggest that physicians rely substantially on experiential pattern recognition and contextual judgment that may not be fully captured in documentation or externalized reasoning traces. This becomes particularly important in the context of AI systems, since many contemporary clinical AI tools are trained primarily on explicit documentation and may therefore fail to capture important implicit aspects of physician cognition.

\paragraph{Documentation Externalization Index}
The Documentation Externalization Index was developed to examine the relationship between physician cognition and formal clinical documentation. This index was calculated by averaging responses regarding the proportion of reasoning physicians consciously articulate and the proportion of reasoning ultimately documented in clinical notes. The purpose of the index was to assess the extent to which reasoning processes are externalized into durable clinical records versus remaining cognitively internal. Higher scores on this index indicate more extensive externalization of reasoning into documentation, whereas lower scores suggest greater separation between cognition and recorded notes.

\paragraph{AI Alignment Index}
The AI Alignment Index was constructed to assess how closely physicians perceive AI-generated documentation to reflect their actual reasoning processes. This index was calculated by averaging ratings of AI-generated note accuracy and alignment between AI-generated notes and physician reasoning. Conceptually, this index captures the degree of representational alignment between physician cognition and AI-mediated clinical representations. This dimension is important because the study argues that the major limitation of current AI systems is not just the predictive accuracy, but also the mismatch between how physicians cognitively organize patient trajectories and how AI systems compress or structure clinical information as well. Lower alignment scores suggest that physicians perceive AI-generated documentation as omitting things like contextual nuance, or other pieces of information that are crucial for clinical reasoning. The index therefore serves as a quantitative proxy for evaluating the representational compatibility between physician cognition and AI-generated outputs.

\paragraph{Importance of Longitudinal AI Support}
Finally, the analysis separately examined ratings related to the Importance of Longitudinal AI Support. Although this measure was retained as an individual dimension rather than incorporated into a larger composite index, it plays a critical interpretive role in the study. This measure assessed the extent to which physicians desired future AI systems capable of supporting longitudinal reasoning tasks (e.g., cross-encounter synthesis, trajectory tracking, unresolved-question monitoring, temporal reconstruction of patient history, etc.). The importance of this variable lies in its ability to connect current dissatisfaction with existing AI tools to physicians’ broader vision for future clinical AI systems. 

\subsection{Results and Discussion}
Physicians reported strongly longitudinal reasoning (Longitudinal Reasoning Index $M = 4.03$, $SD = 0.52$) alongside only moderate alignment between that reasoning and AI-generated notes (AI Alignment Index $M = 2.65$, $SD = 0.74$). The longitudinal pattern was consistent across its component measures: importance of prior encounters ($M = 4.46/5$), extent of trend-based reasoning ($M = 3.74/5$), and frequent referencing, reinterpretation, and retrospective updating of prior encounters. Physicians thus described reasoning through evolving patient trajectories rather than isolated clinical snapshots. Respondents also reported a reliance on active selection rather than exhaustive processing, with mean selectivity in identifying clinically relevant information at $3.77/5$ and confidence in those judgments at $4.08/5$, indicating that reasoning depends on prioritization and contextual interpretation.

Important components of reasoning also remained difficult to verbalize or document. Respondents reported moderate-to-high reliance on intuition ($M = 3.62/5$) together with high reliance on explicit reasoning ($M = 3.95/5$) and moderate difficulty verbalizing their reasoning, yielding a Tacit Reasoning Index of $M = 3.54$ ($SD = 0.71$). Open-ended responses repeatedly cited uncertainty, contextual judgment, evolving trajectories, and alternative diagnoses as elements that remain incompletely externalized, and respondents noted that notes typically capture conclusions rather than full reasoning trajectories. Consistent with this, the Documentation Externalization Index was moderate ($M = 3.21$, $SD = 0.64$), indicating that records often preserve summaries more than the reasoning behind them.

\begin{table}[H]
\centering
\caption{Composite Reasoning and AI Indices}
\begin{tabular}{lcc}
\toprule
\textbf{Index} & \textbf{Mean} & \textbf{SD} \\
\midrule
Longitudinal Reasoning Index & 4.03 & 0.52 \\
Tacit Reasoning Index & 3.54 & 0.71 \\
Documentation Externalization Index & 3.21 & 0.64 \\
AI Alignment Index & 2.65 & 0.74 \\
Importance of Longitudinal AI Support & 3.83 & 0.81 \\
\bottomrule
\end{tabular}
\end{table}

AI use was concentrated on documentation and workflow tasks, such as note drafting, summarization, recall support, and billing, rather than longitudinal reasoning support, and adoption was uneven: 15 physicians reported never using AI-generated notes, 8 rarely, 9 sometimes, 5 often, and 2 always (Figure~1). Physicians were not categorically opposed to AI use. Instead, respondents repeatedly identified only moderate alignment between AI-generated documentation and their actual reasoning processes. Among AI users, mean ratings for AI-generated note accuracy (M = 2.75/5), alignment between notes and physician reasoning (M = 2.54/5), information loss in AI documentation (M = 2.46/5), and distortion of reasoning by AI systems (M = 2.58/5) all suggested partial usefulness alongside substantial perceived limitations. At the same time, physicians expressed substantial interest in future AI systems capable of supporting longitudinal reconstruction and trajectory-based reasoning, with the importance of longitudinal AI support receiving a mean score of 3.83/5 among AI users. Collectively, these findings suggest that the central challenge identified by physicians is a broader representational mismatch between how AI systems organize clinical information and how physicians cognitively construct patient trajectories over time.

Respondents frequently described reasoning as dependent on trend tracking, retrospective reinterpretation, and synthesis across multiple visits rather than single-encounter decision making. These findings support the conceptualization of clinical reasoning as an evolving process of interpretation and revision over time. The survey also demonstrated that substantial portions of physician reasoning remain partially implicit and incompletely externalized in documentation. The Tacit Reasoning Index demonstrated moderate-to-high reliance on intuition and partially implicit judgment (M = 3.54, SD = 0.71). Physicians frequently reported difficulty fully articulating their reasoning processes, particularly under conditions of uncertainty and time pressure.

Documentation practices also reflected incomplete externalization of reasoning processes. The Documentation Externalization Index averaged 3.21 (SD = 0.64), suggesting that clinical notes often capture conclusions and summaries rather than complete reasoning trajectories.

\begin{figure}[H]
\centering
\includegraphics[width=0.8\textwidth]{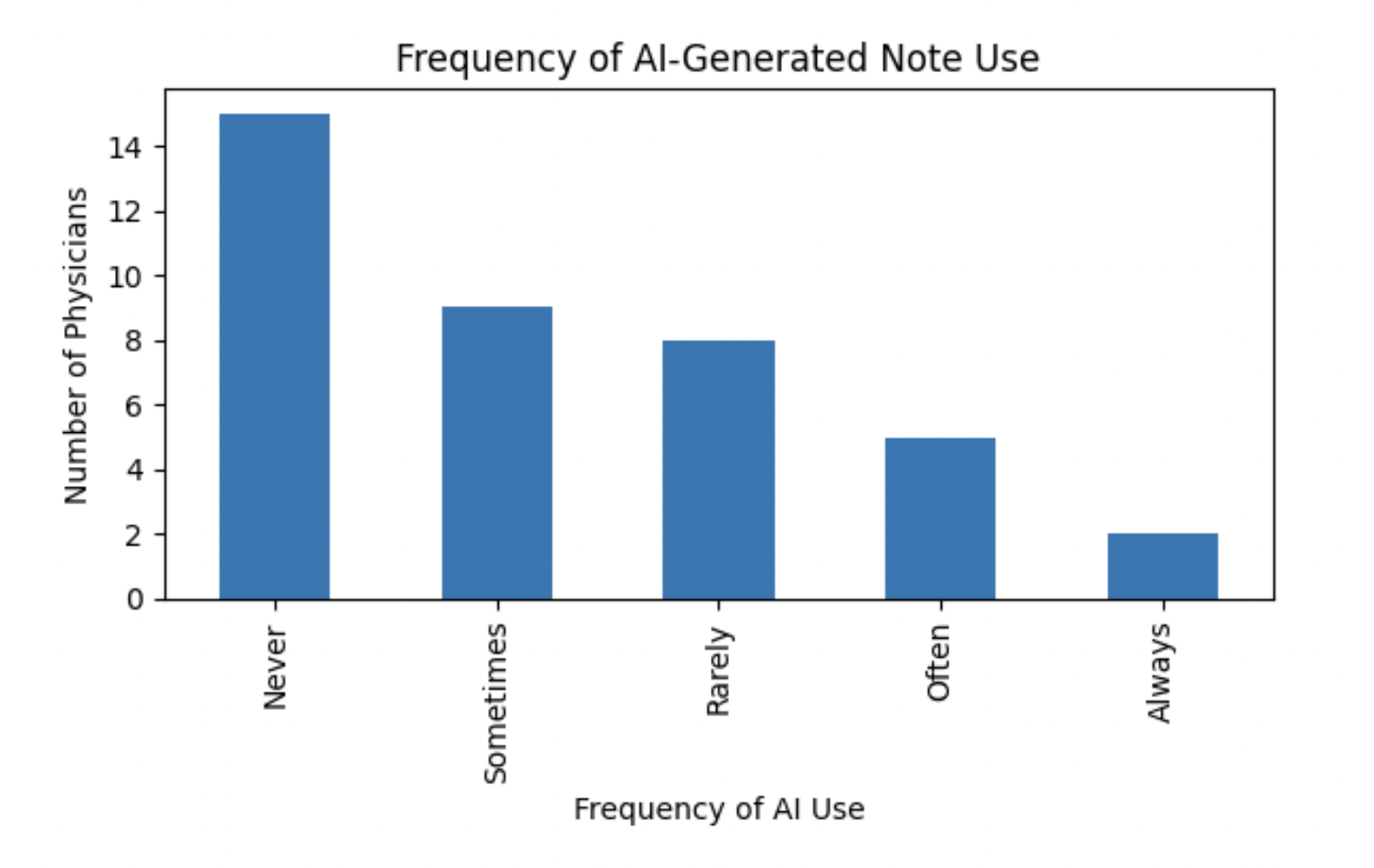}
\caption{Frequency of AI-generated note use by physicians.}
\end{figure}

AI-generated clinical documentation tools were used unevenly across respondents. While some physicians reported regular use of AI-generated note systems, Figure 1 shows that most reported minimal (n = 8) or no (n = 15) adoption. Most AI use was concentrated around encounter-level tasks such as documentation drafting, summarization, and workflow support rather than direct assistance with longitudinal reasoning or diagnostic synthesis.  Respondents frequently identified information loss, compression of temporal context, and omission of uncertainty as limitations of current AI-generated notes.

The AI Alignment Index demonstrated moderate perceived correspondence between AI-generated documentation and physicians’ actual clinical reasoning processes, with an overall mean score of 2.65 (SD = 0.74) out of 5. This suggests that clinicians viewed current AI systems as partially aligned with their reasoning, while still identifying substantial gaps between internal clinical cognition and AI representation. Qualitative responses consistently reinforced this interpretation. Physicians frequently described AI tools as useful for things like summarization, note drafting, recall support, and organization of information, but substantially weaker at capturing uncertainty, contextual judgment, evolving differential diagnoses, longitudinal trajectories, and the weighting of competing possibilities. Several respondents specifically emphasized that AI systems often reproduce conclusions more effectively than the interpretive pathways used to arrive at those conclusions. Collectively, these findings illustrate that contemporary clinical AI systems are perceived primarily as documentation-support technologies rather than systems capable of fully modeling or externalizing the deeper inferential structure of physician reasoning.

The Importance of Longitudinal AI Support Index produced one of the strongest signals in the dataset, with an overall mean score of 3.83 (SD = 0.81) out of 5. This indicates that physicians broadly perceive longitudinal reasoning support as a highly valuable direction for future clinical AI development. Desired capabilities most frequently included tracking changes over time, generating cross-visit summaries, highlighting unresolved issues, representing uncertainty, and synthesizing information across encounters. Importantly, many respondents described their reasoning as fundamentally dependent on temporality: integrating prior encounters, reassessing earlier conclusions in light of new evidence, recognizing evolving patient trajectories, and contextualizing present findings within broader longitudinal histories. The relatively high score for longitudinal AI support aligns closely with the paper’s broader argument that clinical reasoning is inherently cumulative and contextually embedded. 

\begin{figure}[H]
\centering
\includegraphics[width=1\textwidth]{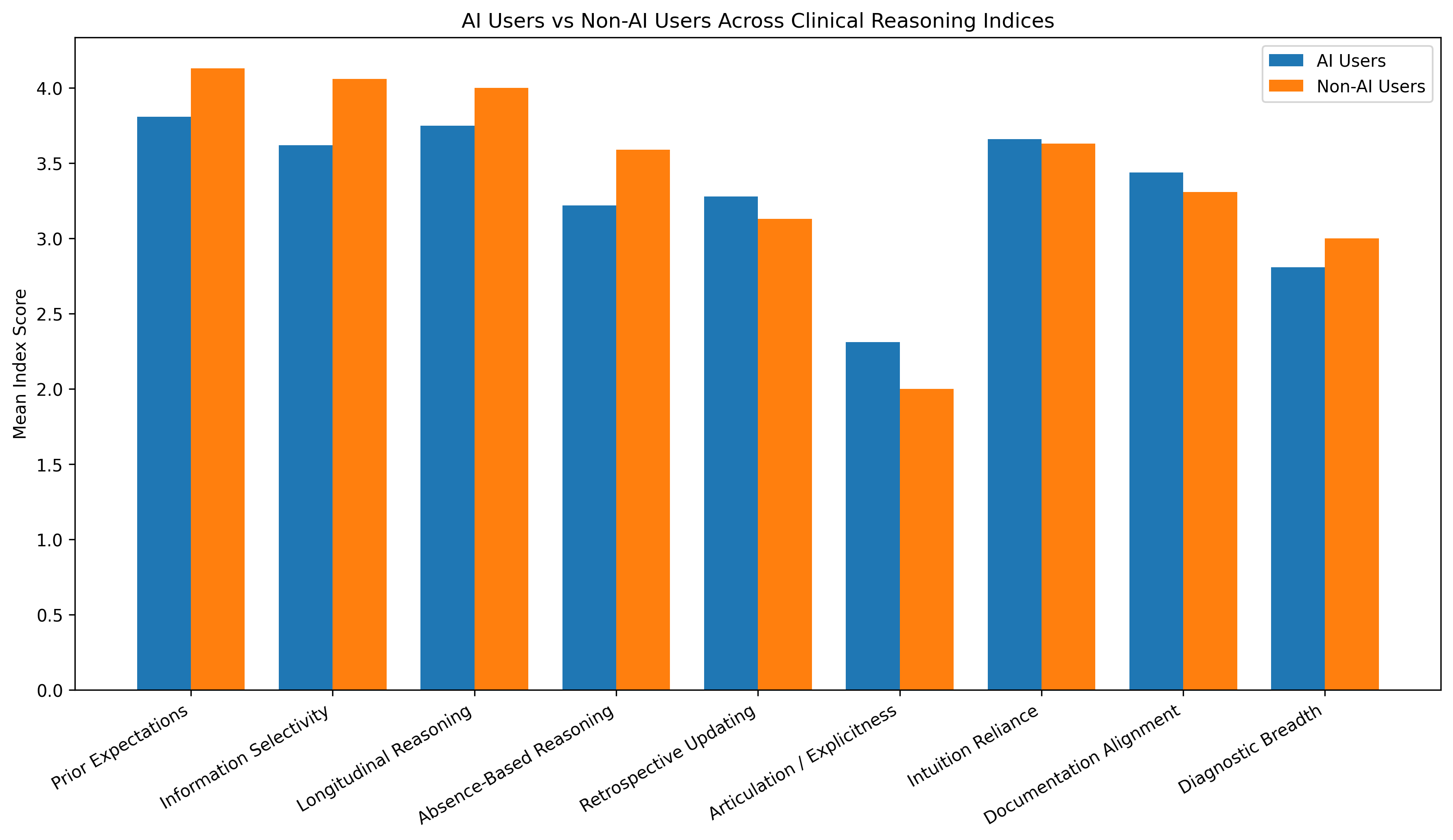}
\caption{Indices and other results by AI users vs. non-AI users.}
\end{figure}

Figure 2 above shows the comparisons between physicians who used AI-generated documentation systems and those who did not, revealing broadly similar reasoning structures across groups. Both AI users and non-users demonstrated high levels of longitudinal reasoning and moderate reliance on tacit judgment processes. Non-AI users demonstrated somewhat higher scores in prior expectation formation, information selectivity, longitudinal trend tracking, absence-based reasoning, and diagnostic breadth. Collectively, these indices reflect reasoning processes that rely heavily on accumulated contextual knowledge across encounters.

By contrast, AI users scored somewhat higher on articulation/explicitness and documentation alignment measures. This may indicate that interaction with AI systems encourages clinicians to externalize or structure parts of their reasoning process in more explicit ways. However, the relatively small differences between groups, especially in intuition reliance, suggest that AI use does not fundamentally replace intuitive or experience-based reasoning. Instead, physicians appear to integrate AI tools into existing cognitive workflows while continuing to rely on tacit clinical judgment.

Importantly, the pattern of results aligns with many of the qualitative responses in the dataset. Physicians repeatedly described AI systems as effective for summarization and recall, while simultaneously emphasizing that these tools struggle to capture longitudinal trajectories, contextual nuance, uncertainty, evolving differential diagnoses, and specialized interpretive reasoning. Several respondents also noted that AI systems often fail to appropriately weight information or incorporate the broader temporal narrative of the patient. In their open-ended answers, AI users reported slightly greater interest in future longitudinal AI support capabilities, including cross-encounter synthesis and trajectory tracking.

These findings support the idea that current clinical AI systems primarily operate as documentation and information management tools rather than as systems capable of reproducing the deeper longitudinal and interpretive dimensions of clinical reasoning. The results reinforce the paper’s central claim that clinical reasoning is not reducible to static documentation outputs or isolated diagnostic predictions. Instead, reasoning emerges across time through contextual and ongoing synthesis of patient trajectories, which does not appear to be represented in present AI workflows.

\section{Study 2: Interviews}
\subsection{Methods}
\subsubsection{Design}
The interview component is designed to capture clinical reasoning as a dynamic, context-dependent process. Using case-based prompts, participants are asked to reconstruct recent challenging patient encounters, walking through their reasoning from initial presentation to diagnosis and management. This narrative format enables the elicitation of sequential reasoning traces, including hypothesis generation, prioritization, revision, and abandonment over time, as well as the identification of moments where things were uncertain. Follow-up probes target specific cognitive processes as well. Additional questions focus on longitudinal reasoning, prompting participants to describe how they track patient trajectories and update prior interpretations. The interviews also elicit metacognitive aspects of reasoning, including the roles of intuition and calibration under time pressure and cognitive load.

A parallel set of interview questions examines physicians’ interactions with AI tools, including both formally integrated systems and informal tools such as large language models. Participants provide concrete examples of AI use within their workflow, describing when AI outputs are helpful versus misleading. These accounts are used to analyze how AI systems intersect with different stages of reasoning, as well as how clinicians calibrate trust in AI outputs.

\subsubsection{Analysis}
The interview component of the study was analyzed using a hybrid deductive–inductive coding framework designed to characterize how physicians describe clinical reasoning in real-world practice and how current AI systems intersect with these reasoning processes. Qualitative data from interviews are analyzed using a coding framework grounded in established constructs from cognitive science and clinical decision-making, including hypothesis generation and revision, temporal inference, uncertainty management, attention and salience filtering, and metacognitive monitoring. Coding is iterative, combining theory-driven categories with inductive refinement to capture emergent patterns. Particular attention is paid to the sequential organization of reasoning and how different processes are coordinated over time. The resulting framework conceptualized reasoning as involving multiple interacting forms of decision making and synthesis. 

\begin{table}[H]
\centering
\caption{Primary Qualitative Coding Categories}
\begin{tabular}{p{4cm} p{10cm}}
\toprule
\textbf{Code} & \textbf{Description} \\
\midrule
Longitudinal Reasoning & Integration of information across encounters over time \\
Trajectory Tracking & Monitoring evolving symptoms, labs, treatments, or disease progression \\
Hypothesis Revision & Updating diagnostic interpretations with new evidence \\
Absence Reasoning & Diagnostic meaning derived from missing findings or lack of expected change \\
Salience Filtering & Selective prioritization of clinically meaningful information \\
Tacit Reasoning & Intuitive or difficult-to-articulate judgment processes \\
Reconstruction Burden & Manual synthesis across fragmented records and systems \\
Representational Mismatch & Disconnect between AI-generated outputs and physician cognition \\
AI Workflow Support & AI use for summarization, drafting, or clerical support \\
Desired AI Capabilities & Physician requests for future longitudinal or uncertainty-aware AI systems \\
\bottomrule
\end{tabular}
\end{table}

Each transcript was coded for the presence and recurrence of several reasoning dimensions. These are shown in Table 2 above.  Additional codes targeted physicians’ interactions with AI systems, including AI-supported documentation, trust calibration, representational mismatch, workflow integration, perceived information loss, and desired future AI capabilities.

Longitudinal reasoning was coded whenever physicians described integrating information across multiple encounters or reconstructing patient trajectories over time. Trajectory tracking captured explicit discussion of monitoring changes in symptoms, laboratory values, treatments, or patient status and appearance over time. Hypothesis revision was coded when physicians described changing diagnostic interpretations in response to evolving evidence. Absence reasoning refers to situations where the lack of expected findings or lack of improvement becomes diagnostically meaningful. Salience filtering refers to the selective prioritization of certain clinical details over others. Tacit reasoning captured moments where physicians described intuition, experiential pattern recognition, or difficulty fully articulating how a judgment was reached.

The AI-related coding structure examined whether physicians used AI systems, as well as how physicians cognitively positioned these systems relative to their own reasoning processes. AI-supported documentation was coded when physicians described AI primarily as a summarization, drafting, or workflow tool. Representational mismatch captured moments where physicians explicitly described disconnects between AI-generated outputs and their clinical reasoning. Desired future AI support captured physician descriptions of AI systems that would better align with their own cognitive processes.

\subsection{Interview Results}
Across the four interviews, several codes appeared with particularly high frequency. Longitudinal reasoning, trajectory tracking, contextual synthesis, and reconstruction burden emerged in nearly every interview and often formed the organizational backbone of physicians’ descriptions of patient care. Tacit reasoning and salience filtering were also highly recurrent, particularly when physicians described how they prioritize information or recognize concerning patient patterns before objective findings fully emerge. AI-related discussions consistently emphasized workflow assistance and documentation support rather than autonomous reasoning replacement. At the same time, physicians repeatedly identified limitations involving temporal or contextual compression and inability of current systems to adequately model longitudinal patient evolution.
  
\subsubsection{Interview 1: Senior Pulmonary and Critical Care Physician}
The first interview involved a senior pulmonary and critical care physician with decades of experience spanning inpatient pulmonary medicine, critical care, pulmonary rehabilitation, and teaching services across multiple hospital systems. The interview was heavily characterized by longitudinal reasoning and tacit clinical judgment developed through long-term clinical experience. Compared to the other interviews, this participant demonstrated the strongest emphasis on temporally extended interpretation and experiential pattern recognition.

One of the most recurrent themes throughout the interview was the idea that patient understanding emerges gradually through longitudinal reconstruction rather than through isolated diagnostic moments. The physician repeatedly described needing to integrate fragmented clinical information across admissions, outpatient follow-up, prior imaging, laboratory trends, medication changes, and evolving symptom patterns. Clinical reasoning was consistently framed not as selecting a diagnosis from a fixed list of possibilities, but as continuously updating an evolving representation of the patient over time. Trajectory tracking appeared throughout the interview, particularly in the discussion of chronic pulmonary disease.

The participant also repeatedly emphasized the importance of contextual synthesis and selective attention. Rather than attempting to process every available piece of information equally, the physician described prioritizing specific findings based expected disease progression or subtle deviations from trajectory. Salience filtering emerged as a dominant coding category in this interview. The physician frequently described identifying clinically meaningful information not because it was objectively the largest abnormality, but because it violated an expected trajectory or did not fit with the overall clinical picture.

Tacit reasoning and intuition were especially prominent in this interview. The participant frequently described situations where concern about a patient emerged before a fully explicit explanation could be articulated. This often involved recognizing subtle combinations of appearance, pacing, respiratory effort, trajectory shifts, or incomplete symptom clusters that collectively generated concern despite limited objective evidence. The physician repeatedly acknowledged that many of these judgments were difficult to formally externalize or document, reinforcing the distinction between clinical cognition and formal documentation systems.

The physician’s discussion of AI systems was notably pragmatic rather than either strongly enthusiastic or dismissive. AI tools were primarily described as useful for documentation efficiency and the reduction of clerical burden. However, the participant repeatedly emphasized that current systems do not adequately capture the longitudinal and interpretive structure of clinical reasoning. Representational mismatch emerged strongly in this interview, particularly around compression of temporal nuance and omission of evolving uncertainty. The physician expressed concern that AI-generated summaries may create overly simplified representations of clinically complex trajectories.

One particularly important theme involved the distinction between documentation support and reasoning support. The participant repeatedly emphasized that while AI may help reduce documentation workload, the actual work of reconstructing patient trajectories remains fundamentally physician-driven. Desired future AI capabilities included systems being able to learn a particular physician's clinical reasoning process over time and provide information based on the unique physician.

\subsubsection{Interview 2: Cardiology Fellow and Internal Medicine Physician}
The second interview involved a cardiology fellow trained in internal medicine who worked across multiple institutional systems, including Veterans Affairs settings, cancer-center specialty care, and general hospital environments where some institutional systems used different EHR infrastructures from each other. This interview placed particularly strong emphasis on fragmentation across systems and the cognitive labor involved in integrating information across incompatible documentation structures.

Longitudinal reasoning and reconstruction burden again emerged as highly recurrent coding categories. The participant described clinical reasoning as heavily dependent on reconstructing patient trajectories across fragmented EHR environments, often requiring manual integration of outpatient records, inpatient consults, specialist notes, imaging studies, laboratory trends, and prior procedural history. The physician repeatedly emphasized that understanding a patient frequently requires synthesizing information distributed across multiple systems, as opposed to relying on any single encounter note.

A major theme throughout this interview involved representational fragmentation within the EHR itself. The participant described the EHR less as a unified cognitive support system and more as a dispersed archive requiring active physician reconstruction. Clinical reasoning, therefore, depended heavily on selective retrieval and retrospective synthesis. Trajectory tracking appeared repeatedly in discussions of cardiovascular disease progression, treatment response, medication adjustment, and evolving cardiac function over time of patients.

Hypothesis revision and retrospective reinterpretation were also especially prominent in this interview. The physician frequently described cases where evolving laboratory values or treatment responses forced reconsideration of earlier interpretations. The interview portrayed reasoning as iterative and dynamically revisable. Several examples involved reinterpretation of earlier assumptions once additional longitudinal context became available.

This participant also described extensive use of salience filtering under conditions of information overload. The physician repeatedly emphasized that modern EHR systems contain enormous quantities of information, much of which is not equally relevant for immediate reasoning tasks. As a result, physicians develop strategies for rapidly identifying clinically meaningful details while filtering out low-priority or redundant material. This selective attention process was repeatedly described as experience-dependent and difficult to formalize.

The AI-related discussion in this interview was highly nuanced. The participant described AI systems as potentially valuable for information organization and summarization, but repeatedly emphasized that current tools remain largely encounter-centered and insufficiently trajectory-aware. The physician expressed concern that AI-generated notes may flatten temporal complexity and obscure unresolved uncertainty by producing overly coherent summaries that fail to preserve ambiguity or evolving interpretation. Importantly, the physician did not frame AI limitations primarily as problems of factual inaccuracy. Instead, the participant repeatedly described a mismatch between how AI systems organize information and how clinicians cognitively reconstruct patient trajectories. This distinction became one of the strongest conceptual findings across the interviews. The physician expressed substantial interest in future AI systems capable of tracking unresolved diagnostic questions or patient information across institutions and encounters.

\subsubsection{Interview 3: Early-Career Internal Medicine Resident Interested in Cardiology}
The third interview involved an early-career internal medicine resident with an emerging specialization interest in cardiology. This interview provided especially useful insight into how developing clinicians learn to structure reasoning under conditions of uncertainty, varying degrees of supervision, cognitive overload, and evolving expertise.

A dominant theme throughout the interview was the active construction of reasoning frameworks during training. The participant frequently described reasoning as a process of learning how to organize or prioritize information and construct coherent patient narratives rather than just memorizing diagnoses. Hypothesis generation and iterative refinement appeared frequently throughout the interview.

Trajectory tracking and longitudinal reasoning were again very recurrent themes, although often discussed in terms of learning how to identify clinically meaningful temporal changes rather than relying on deeply established experiential intuition. The resident repeatedly emphasized the importance of comparing current presentations against prior baselines and monitoring whether patients followed expected trajectories.

One especially important theme involved uncertainty management. The participant repeatedly described uncertainty as a constant feature of clinical reasoning rather than an occasional exception. Instead of expecting immediate certainty, the resident framed reasoning as a process of probabilistic updating in response to new evidence, attending feedback, imaging results, laboratory changes, and patient evolution over time. Several examples involved maintaining multiple competing hypotheses simultaneously while gathering additional information.

Tacit reasoning appeared differently in this interview compared to the more senior physicians. Rather than describing deeply internalized intuition, the participant often described developing emerging pattern recognition capacities while still relying heavily on explicit frameworks and structured reasoning approaches. This contrast may represent an important developmental dimension of clinical expertise, where tacit reasoning becomes increasingly internalized with experience. The participant also discussed the substantial cognitive burden associated with reconstructing patient histories from fragmented notes and documentation systems. Even as a trainee, the resident repeatedly described spending large amounts of time manually synthesizing trajectories across multiple encounters or admissions.

The AI-related discussion was somewhat more optimistic than in the senior physician interviews, although still cautious. The participant described AI tools as potentially useful for recall support and reduction of documentation burden. However, the resident also repeatedly emphasized that AI systems currently struggle with prioritization of clinically meaningful information. The physician expressed concern that trainees might become overly reliant on simplified summaries without fully reconstructing underlying patient trajectories themselves. Desired future AI capabilities included improved longitudinal synthesis and tools capable of helping clinicians reconstruct evolving patient narratives more efficiently. Importantly, the participant repeatedly emphasized that AI should support rather than replace physician reasoning.

\subsubsection{Interview 4: Internal Medicine Resident}
The fourth interview involved a second-year internal medicine resident working in inpatient and residency training environments. Compared to the prior interviews, this interview placed particularly strong emphasis on workflow intensity under time pressure, as well as the tension between cognitive reasoning and documentation demands. Longitudinal reasoning and trajectory reconstruction again emerged as central coding categories. The participant repeatedly described needing to understand how patients changed over time, particularly during prolonged admissions and and chronic disease management. 

A major theme throughout the interview involved information overload and prioritization under workflow pressure. Salience filtering appeared extremely frequently, with the participant repeatedly describing the necessity of rapidly identifying the most clinically meaningful information from very large volumes of documentations or lab results. The physician emphasized that this filtering process often depends heavily on contextual interpretation and prior experience.

Absence reasoning and counterfactual interpretation also appeared prominently in this interview. The participant frequently described reasoning through what was missing rather than only what was present. Examples included lack of expected improvement after treatment or missing findings that would normally support a particular diagnosis. These moments often triggered reconsideration of earlier assumptions and the generation of alternative hypotheses.

The interview also strongly reinforced the distinction between cognition and documentation. The participant repeatedly described clinical notes as compressed artifacts shaped by workflow constraints or billing structures rather than representations of reasoning itself. Several examples highlighted how physicians often retain substantial portions of reasoning internally while documenting only final interpretations or clinically actionable summaries.

AI use was again described primarily in relation to documentation support and reduction of clerical burden. The participant expressed appreciation for AI tools that reduce repetitive writing tasks but emphasized that current systems fail to capture uncertainty or longitudinal nuance. The physician expressed concern that AI-generated summaries may appear superficially coherent while obscuring important contextual subtleties or unresolved diagnostic ambiguity. One especially important theme involved trust calibration. The participant described needing to actively evaluate whether AI-generated summaries accurately reflected the patient trajectory, or whether important contextual information had been omitted. Desired future AI capabilities included systems capable of highlighting trajectory changes, preserving unresolved questions, surfacing contradictions across encounters, and assisting longitudinal patient reconstruction without collapsing uncertainty into oversimplified summaries, in ways that physicians could trust.

\subsubsection{Cross-Interview Discussion}
Across all four interviews, several similar themes emerged. First, clinical reasoning was described as reconstructive and trajectory-based, not encounter-bounded. In other words, physicians emphasized the importance of integrating information across time and reconstructing prior trajectories as being important to their clinical reasoning processes. Second, physicians saw reasoning as partially tacit and incompletely externalized. Across interviews, clinicians said that important aspects of reasoning remain difficult to formally articulate or fully document. Third, all participants described a substantial reconstruction burden associated with fragmented EHR systems and distributed documentation structures. Fourth, AI systems were framed as documentation and workflow tools rather than direct reasoning replacements. Finally, physicians across all interviews expressed a strong interest in future AI systems capable of supporting longitudinal reasoning, such as being able to track trajectories or conduct cross-encounter synthesis. 

Collectively, the interview findings reinforce the central argument of the study: the primary challenge for future clinical AI systems may not just involve improving predictive accuracy, but designing representations that better align with the temporal and partially implicit structure of physician reasoning.

\section{Discussion}

The findings from both the survey and interview components converge on a central conclusion: clinical reasoning is fundamentally longitudinal, reconstructive or interpretive, and only partially externalized within formal clinical documentation systems. Across quantitative and qualitative data, physicians described reasoning as an evolving process of integrating trajectories, revising interpretations, monitoring changes over time, and synthesizing fragmented contextual information across multiple encounters. The survey data demonstrated high levels of reliance on prior encounters, trend tracking, retrospective reinterpretation, and longitudinal integration, while the interviews provided detailed accounts of how these processes unfold in real clinical environments. This finding is consistent with cognitive theories of medical expertise, which suggest that clinicians organize knowledge through illness scripts and context-sensitive representations rather than isolated diagnostic rules \citep{Schmidt1990}. This distinction becomes critically important when considering the design of AI systems, many of which remain organized around encounter-level summarization and not trajectory-level reasoning.

The interview findings deepened and contextualized the survey results by illustrating the mechanisms through which longitudinal reasoning actually occurs in practice. The survey data alone suggested the importance of longitudinal reasoning, but the interviews demonstrated the cognitive labor required to sustain this form of reasoning under real workflow conditions. This reconstruction burden appeared pronounced in environments involving fragmented EHR systems and chronically ill patients with prolonged or recurrent trajectories. The consistency of this finding across physicians of different specialties and training levels suggests that longitudinal reconstruction may represent a foundational feature of clinical reasoning rather than a specialty-specific phenomenon.

A second major finding emerging across both datasets concerns the partially tacit nature of physician cognition. Survey respondents reported moderate-to-high reliance on intuition and substantial difficulty fully verbalizing their reasoning processes, while the interviews repeatedly demonstrated how physicians rely on contextual judgment, experiential pattern recognition, salience filtering, and evolving “gut feelings” that are often difficult to formalize explicitly. Importantly, tacit reasoning reflected compressed forms of expertise developed through repeated exposure to clinical trajectories over time. Senior physicians, in particular, described recognizing concerning patterns before objective findings fully crystallized, often based on subtle contextual deviations or violations of expected trajectories. At the same time, trainees described gradually developing these interpretive capacities while still relying more heavily on explicit frameworks and structured reasoning approaches. Together, these findings suggest that tacit reasoning is not the absence of structure, but instead the progressive internalization of longitudinal and contextual pattern recognition through clinical experience.

The study also revealed a consistent distinction between physician cognition and clinical documentation. Across both the survey and interviews, physicians described documentation as a compressed and incomplete representation of reasoning rather than a comprehensive externalization of cognition itself. Clinical notes were often framed as communication artifacts shaped by workflow constraints and time pressure, or even things like institutional conventions and billing requirements. This distinction is theoretically important because many current AI systems operate primarily on documentation-derived representations of clinical care. If documentation systematically omits things like uncertainty or contextual synthesis, then AI systems trained on these representations may inherit structural limitations regarding what aspects of reasoning remain visible or learnable. In this sense, the study suggests that there may be an important difference between modeling documented medicine and modeling clinical reasoning itself.

The findings regarding AI systems were nuanced and consistent across both methodologies. Physicians did not appear categorically resistant to AI integration, nor did they frame AI primarily as a threat to physician expertise. Instead, most participants described current AI systems as useful but cognitively limited tools that primarily support documentation or identifying medical research. Survey results demonstrated uneven but meaningful adoption of AI-generated documentation systems, while the interviews clarified how physicians position these systems within broader clinical workflows. Importantly, physicians distinguished between documentation support and reasoning support. AI systems were often perceived as capable of accelerating note production or organizing information, but not as systems that adequately model the temporal and reconstructive structure of clinical reasoning itself.

One of the most important findings of the study concerns the idea of representational mismatch between physician cognition and current AI-generated clinical representations. Notably, physicians rarely framed limitations in terms of factual inaccuracy. Instead, concerns centered on how AI systems organize and represent information. This distinction has important implications for the future direction of clinical AI research. A system may produce factually accurate summaries while still failing to preserve the contextually layered structure through which physicians actually reason. This extends prior work on interpretability and clinical explainability by suggesting that clinician-facing AI systems must be evaluated on whether they preserve the representational structure needed for longitudinal clinical reasoning in addition to accuracy \citep{Rudin2019, Tonekaboni2019}.

In terms of desires for future AI systems, physicians described interest in systems that could track unresolved questions over time, synthesize trajectories across encounters, preserve contextual nuance, identify meaningful deviations from expected progression, and assist in the reconstruction of patient narratives across fragmented institutional systems. Importantly, physicians rarely described wanting AI systems that autonomously replace physician reasoning. Instead, participants consistently framed ideal AI systems as cognitive support infrastructures that augment longitudinal synthesis while preserving physician oversight. This aligns with human-AI collaboration research suggesting that effective AI systems should be designed to optimize human-AI team performance rather than maximize standalone model performance alone \citep{Bansal2021, Seeber2020}, suggesting that the future of clinical AI may depend less on replacement-oriented models of automation and more on collaborative systems.

\section{Limitations}

This study has several limitations that should be considered when interpreting the findings on clinical reasoning and AI use. First, physician reasoning is inherently difficult to observe directly. The study relies primarily on interviews and self-reported accounts of clinical decision-making, which necessarily involve retrospective reconstruction rather than real-time observation of cognitive processes. As a result, responses may reflect how physicians conceptualize or narrativize their reasoning after the fact, rather than how reasoning unfolds in practice. Relatedly, recall bias is a potential concern. Physicians may unintentionally rationalize their reasoning when describing past clinical encounters. This may lead to systematic differences between reported reasoning processes and actual in-the-moment decision-making, particularly in complex or high-pressure clinical settings. Future work can mitigate this by synthetically eliciting clinical reasoning with “serial case presentations”, where a case or vignette is presented in segments, and the physician is asked to describe their reasoning at each stage.

A further limitation concerns sampling. The physician sample may not be representative across specialties, institutions, geographic regions, or levels of AI adoption. Clinical reasoning structures and workflows can vary substantially between domains such as emergency medicine, primary care, oncology, psychiatry, intensive care, and other specialties, and the present study may not fully capture this heterogeneity. In addition, differences in institutional EHR infrastructure and availability of AI tools may shape both reasoning practices and reported perceptions of AI. The study is also limited by its reliance on self-reported AI usage. Physicians’ descriptions of how they use AI tools may diverge from their actual workflow behavior, particularly in settings where AI is embedded indirectly into EHR systems or where usage is intermittent and informal. Moreover, because current AI tools vary widely in functionality and integration, physician responses may be shaped more by the specific systems available at their institution than by AI systems in general.

Another limitation is that this study is descriptive and interpretive, not interventional. It characterizes physician reasoning and AI interaction but does not evaluate a newly designed AI system or measure causal effects of AI on reasoning performance. As such, the findings should not be interpreted as evidence of the effectiveness or inefficacy of any particular AI technology. However, this paper calls attention to what future directions AI technologies could take in order to best support clinicians and their work. 

Finally, the survey and interview measures used to operationalize clinical reasoning necessarily represent approximations of complex cognitive and sociotechnical processes. While these measures allow for structured comparison across participants, they cannot fully capture the depth of clinical reasoning as it occurs in real-world practice.

\section{Future Work}

Future research should build on these findings by moving toward more direct and granular observation of clinical reasoning in practice. One key direction is the development and evaluation of AI systems explicitly designed for longitudinal reasoning rather than encounter-level summarization. Current tools often compress clinical history into discrete snapshots, whereas physician reasoning frequently depends on reconstructing trajectories over time. Designing systems that preserve temporal structure and evolving clinical hypotheses would more closely align with real-world cognitive demands.

In parallel, future work should develop evaluation benchmarks that go beyond static accuracy metrics. These benchmarks should assess capabilities such as temporal reasoning, patient trajectory reconstruction, uncertainty representation, counterfactual reasoning, and cross-encounter synthesis, to name a few. Such metrics would better capture whether AI systems support the kinds of reasoning physicians actually perform.

Experimental and quasi-experimental studies are also needed to compare physician reasoning with and without AI-generated longitudinal summaries. These studies could assess whether AI tools reduce cognitive load during complex case reconstruction, particularly in cases involving fragmented records or multi-institutional care histories. Closely related work should examine whether AI support meaningfully changes documentation practices and whether this, in turn, alters the depth or structure of clinical reasoning.

Future research should also incorporate direct workflow observation methods, such as screen-recorded EHR interactions, real-time shadowing, or EHR log analysis. These approaches would allow for comparison between reported reasoning strategies and observed behavior, helping to address limitations of self-report data and retrospective accounts. In addition, future systems could explicitly represent the evolution of diagnostic hypotheses over time, making visible how interpretations of a patient’s condition change across encounters. Such interfaces could support metacognitive awareness and improve coordination in longitudinal care.

Finally, an important open question is how AI-generated documentation affects physician cognition over time. Future studies should investigate whether AI support reduces cognitive burden and frees attention for higher-level reasoning, or whether it risks reducing physician engagement with patient narratives and weakening reflective practice. Understanding this balance will be critical for designing AI systems that enhance rather than inadvertently erode clinical reasoning quality.

\section{Conclusion}
This study examined how physicians reason across longitudinal patient histories and how those reasoning processes align with current AI-assisted clinical documentation systems. Across both quantitative and qualitative findings, a consistent pattern emerged: clinical reasoning is dynamic and cumulative, whereas current AI systems frequently produce static, incomplete representations. The challenge identified by this work is therefore not simply one of model accuracy, but of representational alignment between computational systems and the cognitive processes they are intended to support as well.

These findings have important implications for how clinical AI systems should be evaluated. Current benchmarks understandably emphasize predictive accuracy, summarization quality, documentation completeness, or task-specific performance, but such metrics provide only a partial assessment of a system's utility for supporting physician reasoning. We argue that future evaluation frameworks should also consider whether AI systems preserve longitudinal patient trajectories, represent uncertainty and evolving hypotheses, maintain evidence provenance across encounters, and facilitate retrospective reinterpretation as new clinical information becomes available. Systems with equivalent predictive performance may differ substantially in their ability to support longitudinal reasoning, suggesting that representational alignment constitutes an independent dimension of evaluation.

The findings likewise carry implications for AI system design. Rather than treating patient histories as information to be compressed into isolated predictions or static summaries, future systems may benefit from representing patient state as an evolving temporal process. Designing AI around longitudinal patient state rather than encounter-level outputs represents a shift from optimizing prediction alone toward supporting interpretation and synthesis.

More broadly, this work argues that longitudinal fidelity should be recognized as a fundamental design principle for EHR-integrated AI (see, e.g., \cite{Yi2026LongitudinalEHR}). The extent to which a system preserves the continuity or temporal organization of patient histories is likely to influence its usefulness for complex clinical decision-making independently of predictive performance. 

Finally, these findings have implications for the future of human-AI collaboration in medicine. Rather than replacing physician reasoning, AI systems are likely to function most effectively as cognitive partners that organize and surface clinically meaningful information while leaving interpretation and judgment to clinicians. In this view, AI-generated outputs become another input into an ongoing process of human reasoning rather than a substitute for that reasoning itself. Developing systems that align with this nature of physician cognition may ultimately prove as important as improving algorithmic performance, and may represent the next major challenge for the design and evaluation of clinical AI.

\bibliographystyle{apalike}
\bibliography{sample}

\end{document}